\documentstyle[12pt]{article}
\begin{document}
\renewcommand{\thebibliography}[1]{ {\vspace{5mm}\par \noindent{\bf
References}\par \vspace{2mm}}
\list
 {\arabic{enumi}.}{\settowidth\labelwidth{[#1]}\leftmargin\labelwidth
 \advance\leftmargin\labelsep\addtolength{\topsep}{-4em}
 \usecounter{enumi}}
 \def\newblock{\hskip .11em plus .33em minus .07em}
 \sloppy\clubpenalty4000\widowpenalty4000
 \sfcode`\.=1000\relax \setlength{\itemsep}{-0.4em} }
\newcommand\rf[1]{(\ref{#1})}
\def\nn{\nonumber}
\newcommand{\ft}[2]{{\textstyle\frac{#1}{#2}}}
\newcommand{\eqn}[1]{(\ref{#1})}
\renewcommand{\a}{\alpha}
\renewcommand{\b}{\beta}
\renewcommand{\c}{\gamma}
\renewcommand{\d}{\delta}
\newcommand{\pa}{\partial}
\newcommand{\g}{\gamma}
\newcommand{\G}{\Gamma}
\newcommand{\A}{\Alpha}
\newcommand{\B}{\Beta}
\newcommand{\D}{\Delta}
\newcommand{\e}{\epsilon}
\newcommand{\E}{\Epsilon}
\newcommand{\z}{\zeta}
\newcommand{\Z}{\Zeta}
\newcommand{\K}{\Kappa}
\renewcommand{\l}{\lambda}
\renewcommand{\L}{\Lambda}
\newcommand{\La}{\Lambda}
\newcommand{\m}{\mu}
\newcommand{\M}{\Mu}
\newcommand{\n}{\nu}
\newcommand{\N}{\Nu}
\newcommand{\x}{\chi}
\newcommand{\X}{\Chi}
\newcommand{\p}{\pi}
\newcommand{\R}{\Rho}
\newcommand{\s}{\sigma}
\renewcommand{\S}{\Sigma}
\newcommand{\T}{\Tau}
\newcommand{\y}{\upsilon}
\newcommand{\Y}{\upsilon}
\renewcommand{\O}{\Omega}
\newcommand{\q}{\theta}
\newcommand{\pl}{-i(X^J {\bar F}_J - {\bar X}^J F_J) }
\newcommand{\be}{\begin{equation}}
\newcommand{\ee}{\end{equation}}
\newcommand{\ov}{\overline}
\newcommand{\un}{\underline}
\newcommand{\la}{\langle}\newcommand{\ra}{\rangle}
\newcommand{\bl}{\boldmath} \newcommand{\ds}{\displaystyle}
\newcommand{\nl}{\newline} \newcommand{\wt}{\widetilde}
\newcommand{\wh}{\widehat} \newcommand{\ms}[1]{\mbox{\scriptsize #1}}
\newcommand{\fs}[1]{\mbox{\scriptsize \bf #1}}
\def\beqa{\begin{eqnarray}} \def\eeqa{\end{eqnarray}}
\def\bea{\begin{eqnarray}} \def\eea{\end{eqnarray}}

\thispagestyle{empty}
\begin{center}
\hfill THU-99/15\\
\hfill {\tt hep-th/9906095}

\vspace{3cm}

{\large\bf Modifications of the area law and N=2 supersymmetric black
holes}\\

\vspace{1.4cm}

{\sc Bernard de Wit}\\

\vspace{1.3cm}

{\em Institute for Theoretical Physics, Utrecht University}\\
{\em 3508 TA Utrecht, The Netherlands} \\

\vspace{1.2cm}

\centerline{\bf Abstract}
\vspace{- 4 mm}  \end{center}
\begin{quote}\small
Modifications of the area law are crucial in order to find agreement
with microscopic entropy calculations based on string theory, when
including contributions that are subleading for large charges. The
deviations of the area law are in accord with Wald's proposal for the
entropy based on a Noether charge. We discuss this for the case
of four-dimensional $N=2$ supersymmetric black holes. 
\end{quote}
\vfill

\baselineskip18pt
\noindent  
It is a pleasure and an honour to speak at this conference
and  pay tribute to Francois Englert's scientific achievements. This
meeting brings back so many memories of previous encounters with him
which 
were both enjoyable and scientifically fruitful. I will speak about
some recent work, done in collaboration with Gabriel
Lopes Cardoso and Thomas Mohaupt, on the entropy of $N=2$
supersymmetric black holes \cite{CDWM}.  
The first law of black hole mechanics relates a change in the mass and
angular momentum of a black  
hole to the change in the area of its event horizon. This then leads
to the Bekenstein-Hawking area law, which expresses the
black hole entropy in terms of the horizon area. 
The first law relates global
quantities of a black hole, such as its energy or mass and its
angular momentum which can be determined from the behaviour
of the fields at spatial infinity, to the horizon area which is
defined at the inner boundary of the black hole solution. The
similarity with thermodynamics suggests a possible  
interpretation of the entropy in terms of microstates. Such a 
connection has recently been provided in the context of string
theory \cite{StromVafa}. More specifically, for four-dimensional
extremal black holes 
in the limit of large electric/magnetic charges $Q$, it was found that
the entropy is generically of the form  
\be
{\cal S}\sim \sqrt{Q^4}\,. \label{s=q2}
\ee
This result can be compared to macroscopic results based on
corresponding effective field theories. These theories are complicated
and will depend on many fields and eventually involve higher-derivative 
interactions. As it turns out, the latter interactions are related to
corrections to \eqn{s=q2} that are subleading in the
charges and of order $Q^0$. Such subleading terms have recently been
obtained for IIA string theory and M-theory \cite{MSWV} and full
agreement with the macroscopic description has been established in
\cite{CDWM}.  

How can the results of the
macroscopic description, which is very complicated, yet be so constrained
as to precisely reproduce the results from the counting of
microstates? Before answering this question let me first
characterize the black-hole solutions that I will be discussing. 
They are four-dimensional, static, rotationally symmetric solitonic
interpolations between  
two $N=2$ supersymmetric groundstates: flat Minkowski spacetime at spatial 
infinity and Bertotti-Robinson spacetime at the horizon.  Globally the
solution possesses only a residual $N=1$ 
supersymmetry, so that one is dealing with a BPS state and thus an
extremal black hole. The effective field theory is based on $N=2$
supergravity coupled to a number of vector multiplets whose gauge
fields are 
associated with electric and magnetic charges, denoted by
$q_I$ and $p^I$, respectively. The theory incorporates, in a
systematic fashion, the phenomenon of electric/magnetic duality,
according to which the electric and magnetic charges can be
interchanged and/or rotated and it includes higher-derivative
couplings with among them a certain class of 
terms quadratic in the Riemann tensor. 
 
The crucial ingredient that is responsible for the  remarkable
restrictions on the entropy formulae obtained on the basis of these
complicated effective field theories, is the enhancement to full
supersymmetry at the horizon. However,  
it turns out that in order to obtain agreement for the first
subleading corrections with the counting of
microstates provided by string theory, one is forced to depart from
the Bekenstein-Hawking area law. Instead one must adopt
Wald's proposal for the entropy which ensures the validity of the
first law of black hole mechanics for more generic field
theories. This proposal is based on the existence of a Noether charge 
associated with an isometry evaluated at the corresponding  
Killing horizon \cite{Wald}. When evaluating this current subject to
the field equations, current conservation becomes trivial and the
current takes the form of an improvement term, i.e., the
divergence of an antisymmetric tensor. This antisymmetric tensor,
sometimes called the Noether potential, is a local function of the fields
and of the (arbitrary) gauge transformation parameters. 

In order to elucidate some of this this let me briefly consider a
simple abelian gauge theory, with a gauge-invariant Lagrangian 
depending on the field strength $F_{\m\n}$,  its derivatives $\pa
_\rho F_{\m\n}$,  as well as on matter  
fields $\psi$ and first derivatives thereof. Furthermore we add a
 so-called topological mass term, so that the total Lagrangian takes
 the form
\be
{\cal L}^{\rm total} = {\cal L}^{\rm inv}(F_{\m\n},\pa_\rho F_{\m\n},
 \psi, \nabla_{\!\mu\,}\psi) + c\,\varepsilon^{\m\n\rho} A_\m\,\pa_\n
 A_\rho \,,
\ee
where $\nabla_{\!\mu\,}\psi$ is the covariant derivative of
 $\psi$ and $c$ is some constant. Note that the topological mass term
 restricts us to three 
 spacetime dimensions, but this is not relevant for what
 follows. Under gauge transformations  
\be
\delta_\xi A_\mu(x) = \pa_{\!\mu\,} \xi(x)\,,\qquad
\d_\xi \psi(x) = i \xi(x)\, \psi(x)\,,
\ee
the Lagrangian is not invariant but changes into a total derivative,
\be 
\d_\xi{\cal L}^{\rm total} = \pa_\m N^\m(\phi,\xi) =
 c\,\varepsilon^{\m\n\rho}\,\pa_\m \xi \,\pa_\n A_\rho \,,
\ee
where generically $\phi$ denotes all the fields and $\xi$ denotes the
 transformation parameters. 

Under arbitrary variations $\phi\to\phi+ \d\phi$ we have the standard
decomposition into a surface term and the field equations,  
\be
\d{\cal L}^{\rm total} = \pa_\m\theta^\m(\phi,\d\phi) + \d\phi\cdot
 E(\phi)\,. \label{theta}
\ee
Here $E(\phi)$ denotes the field equations. 
The conserved Noether current is then defined by
\be
J^\m(\phi,\xi) = \theta^\m(\phi,\d_\xi\phi) -  N^\m(\phi,\xi) \,,
\ee
where we observe that, unlike the first term, the second term is not
necessarily proportional to the symmetry variations $\d_\xi\phi$. For
the above example the explicit expression for the current reads,
\bea
 J^\m(\phi,\xi)  &=&  2\, {\cal L}^{\mu\nu}\,\pa_{\nu}\xi -
2\pa_{\rho} {\cal L}^{(\rho,\mu)\nu}\, \pa_{\nu} \xi 
+2 \,{\cal L}^{[\rho,\mu]\nu}\,\pa_{\rho} \pa_{\nu\,}\xi  + {\cal
L}^\m \,\xi \psi 
\nonumber\\
&& 
 -2c\, \varepsilon^{\m\n\rho} \xi\,\pa_\n A_\rho  +c^\prime\,
\varepsilon^{\m\n\rho} \pa_\n(\xi\, A_\rho) \,, \quad{~} \label{Ncurrent}
\eea
where 
${\cal L}^{\mu\nu}$, 
${\cal L}^{\rho, \mu\nu}$ and 
${\cal L}^{\mu}$ denote the derivatives of the action with respect to
$F_{\mu\nu}$, $\pa_{\rho} F_{\mu\nu}$ and $\nabla_{\!\m} \psi$,
respectively. Observe that the Bianchi identity implies ${\cal
L}^{[\rho,\mu\nu]}=0$. I also included an
improvement term parametrized by an arbitrary  
constant $c^\prime$, which reflects an ambiguity in defining
$N^\m$. Its significance will become clear shortly, but 
at this point I  note that for $c^\prime = 2c$ the
current depends only on the derivative of $\xi$ (which equals $\d_\xi
A_\m$). On the other hand, the current is only gauge invariant when we
choose $c^\prime =0$.  

Because \eqn{Ncurrent} is conserved for any function
$\xi$, one argues that, for fields satisfying the field
equations, the current can be written as the divergence of a so-called
Noether potential, 
\be
J^\mu = \pa_\n \,{\cal Q}^{\m\n}\,.
\ee
In the example, the Noether potential ${\cal Q}^{\m\n}$ takes the form
\be
{\cal Q}^{\m\n}(\phi, \xi) = 
2 \,{\cal L}^{\mu\nu}\, \xi - 2\,\pa_{\rho}{\cal
L}^{\rho, \mu\nu} \, \xi +  {\cal L}^{\rho,
\mu\nu}\, \pa_{\rho} \xi + c^\prime\,\varepsilon^{\m\n\rho}
A_\rho\,\xi  \,, 
\ee
and is thus a local function of the fields and of the transformation
parameter $\xi$. Observe that ${\cal Q}_{\m\n}$ does not have to
vanish for field 
configurations that are invariant (in the case at hand, this would
imply $\pa_\m\xi = \xi\psi=0$). 

Integration of the Noether potential over the
boundary of some (spacelike) hypersurface leads to a surface charge,
which, when restricting the gauge transformation parameters to those
that leave the background invariant, is equal to the Noether charge in
the usual 
sense. Under certain conditions this surface charge remains constant
under variations that continuously connect 
solutions of the equations of motion. Here I have in mind a continuous
variety of solutions of the field equations which are left invariant
under a corresponding variety of residual gauge
transformations. Hence, the parameters $\xi$ that characterize the
residual symmetry may change continuously with the solution. Denoting
the combined change of the solution $\phi$ and the symmetry 
parameters $\xi$ by the variation $\hat \d$, one may thus write
\be
\hat\d \Big(\d_\xi\phi\Big) = 0\,.
\ee

Now suppose that the Noether current can be written as a function of
$\phi$ and $\d_\xi\phi$. In that case one knows that $\hat\d
J(\phi,\d_\xi\phi)$ remains proportional to $\d_\xi\phi$ and must
therefore 
vanish for the symmetric configurations. Consequently 
$\hat\d{\cal Q}^{\m\n} (\phi,\xi)$ must vanish up to a closed form,
$\pa_\rho \omega^{[\m\n\rho]}$, so that the surface charge obtained by
integration over a Cauchy surface $C$ with volume element ${\rm
d}\Omega^\m$, 
\be
\int_C \, {\rm d}\Omega_\m\,J^\m(\phi,\d_\xi\phi) = \oint_{\pa C}
\,{\rm d}\Sigma_{\m\n} \,{\cal Q}^{\m\n}(\phi,\xi) \,,
\ee
is constant under the variations induced by $\hat\d$. In the example
this situation is realized by choosing $c^\prime = 2 c$, because in
this case the $\xi$-dependence of the current is fully
captured in $\d_\xi\phi$. Observe, however, that
the integrand on the right-hand side is in principle nonvanishing and
nonconstant, so that the constancy of the total surface charge
represents a nontrivial result. 

Let me now return to general relativity and try to apply the same
arguments. Here the gauge transformations are represented by the
diffeomophisms and the 
residual gauge symmetries correspond to certain Killing
vectors. The Lagrangian is not invariant but transforms as a density,
which implies that $N^\m(\phi,\xi) \propto \xi^\m\,{\cal L}$. Under
field variations the Lagrangian will vary into the quantity
$\theta^\m(\phi,\d\phi)$ defined in \eqn{theta}, so that the
variations of the Noether charge will show a 
certain systematics which will reflect itself in the final result. 
When the Lagrangian depends arbitrarily on the Riemann tensor
$R_{\m\n\rho\sigma}$ (but not on its derivatives) and on matter fields
and their first-order derivatives, one can show that the Noether
potential takes the form 
\be
{\cal Q}^{\mu\n}(\phi, \xi) = -2 {\cal L}^{\m\n\rho\sigma}
\,\nabla_{\!\rho\,} \xi_\sigma + \cdots\,, \label{noetgrav}
\ee
where the extra terms depend on $\xi^\m$ but not on derivatives
thereof and ${\cal L}^{\mu\nu\rho\sigma}$ is defined by $\pa{\cal L}/ \pa
R_{\mu\nu\rho\sigma}$. Observe that 
${\cal L}^{\mu\nu\rho\sigma}$ is antisymmetric in $\scriptstyle[\m\n]$ and in
$\scriptstyle[\rho\sigma]$ and it is symmetric under pair exchange and
satisfies the cyclicity property, ${\cal
L}^{\mu\nu\rho\sigma} ={\cal L}^{\rho\sigma\mu\nu}=-2\,{\cal
L}^{\rho[\mu\nu]\sigma}$.  

As was shown by Wald \cite{Wald} one can employ the Noether charge in
order to find 
generalized definitions of the black hole entropy that ensure the
validity of the first law of black hole mechanics. The
crucial observation \cite{Wald} is that there exists a Hamiltonian,
whose change under a variation of the fields (Wald assumes that the
vector field $\xi^\m$ associated with the diffeomorphism  remains
constant) can be expressed in terms 
of the corresponding change of the Noether charge and takes the
following form 
\be
\d H=
 \d\Big (\int_C {\rm d}\Omega_\m \,J^\mu(\phi,\xi)  \Big )  -
2\,\int_C {\rm d} \Omega_\m
\, \nabla_{\!\n\,} \Big( \xi^{[\m}\,\theta^{\n]}(\phi,\d\phi)\Big)
\,. 
\label{variation1} 
\ee
Here the second term is induced by the variation of the $N^\m$
component of the current, which in this case is associated with a
particular timelike Killing vector field $\xi^\m$ whose Killing
horizon coincides with the black hole event horizon. The quantity $H$
can be associated with a Hamiltonian that governs the evolution along
the integral timelike lines of $\xi^\m$. In order that this
Hamiltonian exists, the second term in \eqn{variation1} should be
expressible as the variation of some other term.
Whenever $\xi^\m$ is a Killing vector of the background solution and
$\d\phi$ connects different solution of the equations of motion,  
$\delta H$ will vanish.  On the other hand, when 
imposing the equations of motion, \eqn{variation1} takes the form of
surface integrals over the boundary $\pa C$ with surface element
${\rm d}\Sigma_{\m\n}$,
\be 
\d H= \int_{\pa C} {\rm d}\Sigma_{\m\n} \, \Big(\delta 
{\cal Q}^{\m\n}(\phi,\xi) -  \xi^\m\,\theta^\n(\phi,\d\phi) +
\xi^\n \,\theta^\m(\phi,\d\phi) \Big) \,.
\label{deltah}
\ee
Here it is important that the last two terms are proportional to the
Killing vector and not to its derivatives. I already mentioned that one
assumes that these terms can be rewritten (at least locally)
as variations. Then, up to a 
proportionality factor, the resulting variations of the surface
integrals at infinity correspond to the mass and angular 
momentum variations that one has in the first law, while the surface
integral at the horizon defines the variation of the entropy. Hence
the mass, the angular momentum and the entropy are surface charges derived
from the same current. 
The entropy takes the form of an integral over the Noether potential
\be
{\cal S} =  -\pi \, \int_{\Sigma_{\rm hor}} {\cal Q}^{\mu \nu} \;
\epsilon_{\mu \nu}\; \Big\vert_{\xi^\m=0,\; \nabla_{[\m\,}\xi_{\n]}=
\e_{\m\n}}\;.  
\label{entronoet}
\ee
Here $\Sigma_{\rm hor}$ denotes a spacelike cross section of the
Killing horizon (which usually has the topology of $S^2$) and we
have used ${\rm d}\Sigma_{\m\n} = 
\e_{\m\n}\,\sqrt{h} \,{\rm d}^2x$. Furthermore  
$\epsilon_{\mu \nu}$ denotes the binormal spanned by two lighlike
vectors at the horizon, which is normalized according to
$\e_{\m\n}\e^{\m\n}=-2$. To appreciate the two conditions, 
$\xi^\m=0$ and $\nabla_{[\m\,}\xi_{\n]}=\e_{\m\n}$, we first note that 
the Noether potential ${\cal Q}^{\m\n}$ can generally be
decomposed according to 
${\cal Q}^{\mu \nu} = N^{\mu \nu \rho} \, \xi_{\rho} + 
Y^{\mu \nu \rho \sigma} \nabla_{[\rho\,}\xi_{\sigma]}$.
This is so, because the derivatives of a Killing vector field are not
independent: the first derivative is proportional to the curl owing to
the Killing condition whereas higher derivatives 
are related to lower ones according to the general identity 
$\nabla_{\!\mu} \nabla_{\!\nu\,} \xi_{\rho} = 
 R_{\nu \rho \mu}{}^\sigma \xi_{\sigma}$. Obviously the condition on
the curl of the Killing vector just sets its normalization and that of
the corresponding Noether potential. In many cases the
terms proportional to $\xi^\m$ cannot 
contribute for symmetry reasons. Both these
conditions are subtle for extremal black holes, because the
surface gravity vanishes at the horizon. We leave these issues aside
here and refer to the literature but we stress that the definition  
\eqn{entronoet} applies to both extremal and nonextremal black holes. 

Obviously, the constancy of the surface charge implied by the
vanishing of the left-hand side of \eqn{deltah} then ensures the
validity of the first law. The normalization in \eqn{entronoet} has
been chosen such 
that, with our conventions, we reproduce the Bekenstein-Hawking area
law for static black holes in general relativity. 
In the presence of higher-derivative terms, the entropy
of a static black hole solution will in general not any longer be
given by the area.  When the Lagrangian
depends on the Riemann curvature (but not on its derivatives) 
and on matter fields and their first derivatives, one can
make use of \eqn{noetgrav} so that the entropy of the static black
hole equals 
\be
{\cal S} = 2\pi \int_{\Sigma_{\rm hor}} \;  {\cal L}^{\mu \nu \rho
\sigma} \; \epsilon_{\mu \nu}  \;  \epsilon_{\rho  \sigma} \;.
\label{graventro}
\ee
This is the result I will be using for the supersymmetric black holes.

Let me now turn to the supergravity Lagrangians that give rise to
these extremal black holes, which are based on the coupling of $n$
vector multiplets to $N=2$ 
supergravity. In general they contain various other couplings, such as
those associated with hypermultiplets, which, however, play only a
limited role in the following and will be omitted.
The construction of the coupling of vector multiplets to $N=2$
supergravity utilizes the so-called superconformal multiplet calculus
which enables one to straightforwardly include the 
interactions proportional to the square of the Riemann tensor (the
appropriate references can be found in \cite{CDWM}). 
In the superconformal framework, there is a multiplet, the so-called
Weyl multiplet, which comprises the gravitational degrees of freedom, namely
the graviton, two gravitini as well as various other superconformal
gauge fields and also some auxiliary fields.  One of these
auxiliary fields is an anti-selfdual Lorentz tensor field $T^{ab\, i j}$,
where $i,j=1,2$ denote chiral $SU(2)$ indices. 
The field strengths corresponding to the various gauge fields in the Weyl
multiplet reside in a so-called reduced chiral multiplet, denoted
by $W^{ab\, i j }$, from which one then constructs the unreduced
chiral multiplet $W^2 = (W^{ab\,i j } \varepsilon_{ij})^2$.
The lowest component field of $W^2$ is equal to ${\hat A} = 
(T^{ab\, i j } \varepsilon_{ij})^2$.

In addition, there are $n + 1$ abelian
vector multiplets labelled by an index $I = 0,
\dots, n$. The extra vector multiplet is required to provide the
graviphoton field of supergravity.  Each vector multiplet
contains a complex scalar field $X^I$, a vector gauge field $W_{\mu}^I$
with field strength $F_{\mu \nu}^I$, as well as a doublet of gaugini
and a triplet of auxiliary scalar fields.  The couplings of these
$n+1$ vector multiplets 
to the Weyl multiplet are encoded in a holomorphic function $F(X^I,
{\hat A})$, which is homogenous of degree two and thus satisfies $X^I
F_I + 2 {\hat A} F_{\hat A} = 2 F$, where $F_I =  
\partial F/\pa {X^I}$, $F_{\hat A} = \partial F/\pa {\hat A}$.

The field equations of the vector multiplets are subject to 
equivalence transformations corresponding to electric-magnetic
duality, which do not involve the fields of the Weyl multiplet.
These equivalence transformations constitute 
${\rm SP}(2n+2;{\rm \bf Z})$
transformations.  Two complex $(2n+2)$-component vectors can now be
defined which transform linearly under 
${\rm SP}(2n+2;{\rm \bf Z})$ transformations,
namely
\be
V =  \left( \begin{array}{c} X^I \\ F_J(X,{\hat A}) \\ \end{array} \right)
\;\;\;\;\; {\rm and} \;\;\;\;\;
 \left( \begin{array}{c} F_{\mu \nu}^{\pm I} \\ 
G^\pm_{\mu \nu J}  \\ \end{array} \right)
\;\;, 
\ee
where $(F_{\mu \nu}^{\pm I},G^\pm_{\mu \nu J})$ denotes the
(anti-)selfdual part 
of $(F_{\mu \nu}^{I},G_{\mu \nu J})$.  The field strength $G^\pm_{\mu \nu J}$
is defined by the variation of the action with respect to 
$F^{\pm J}_{\mu \nu}$. 
By integrating the gauge fields over two-dimensional
surfaces enclosing their sources, it is possible to associate
to $(F_{\mu \nu}^{+I},G^+_{\mu \nu J})$ a symplectic vector $(p^I,q_J)$
comprising the 
magnetic and electric
charges. It is then possible to construct a complex quantity $Z$
out of the charges and of $V$ which is invariant under symplectic 
transformations, as follows, 
\be
Z =  e^{ {\cal K}/2} \, (p^I F_I (X,{\hat A}) - q_I X^I) \;,
\label{z}
\ee
where  $e^{-\cal K} = i [\bar{X}^I F_I (X,{\hat A}) - 
\bar{F}_I ({\bar X}, {\bar {\hat A}}) X^I ]$. 

The associated (Wilsonian) effective Lagrangian
describing the coupling of these vector multiplets
to supergravity is complicated. We only display those terms
which will be relevant for the computation of the entropy of
a static supersymmetric black hole,
\be
{8 \pi \cal L} = - \ft{1}{2} e^{-{\cal K}}R 
+ \ft12( {i} F_{\hat{A}} \,\hat{C} +
\mbox{h.c.}) + \cdots \,,
\label{Paction}
\ee
where ${\hat C} = 64 \,C^{- \mu \nu \rho \sigma } 
C^-_{ \mu \nu \rho \sigma}
+ 16 \,\varepsilon_{ij}\, T^{\mu \nu i j} f_{\mu}{}^{\rho\,} T_{\rho \nu k l}
\,\varepsilon^{kl} + \cdots \,$.  Here $C^-_{ \mu \nu \rho \sigma}$
denotes the anti-selfdual 
part of the Weyl tensor $C_{ \mu \nu \rho \sigma}$, and
$f_{\mu}{}^{\nu} = \ft{1}{2} R_{\mu}{}^{\nu} - \ft{1}{12} R \,
\delta_{\mu}{}^{\nu}+ \cdots$.  
Eventually we set
$e^{-{\cal K}} =1$ 
in order to obtain a properly normalized Einstein-Hilbert term.
We note that the Lagrangian contains $C_{ \mu \nu \rho
\sigma}^2$-terms, but no terms 
involving derivatives of the Riemann curvature tensor.  By expanding the
holomorphic function $F(X, {\hat A})$ in powers of ${\hat A}$, 
$F(X, {\hat A}) =\sum_{g=0}^{\infty} F^{(g)}(X) \,{\hat A}^g$, we see that
the Lagrangian (\ref{Paction}) contains an infinite set of 
curvature terms of the type $C^2 ( T^{\mu \nu i j} \varepsilon^{ij}
){}^{2g-2}$  (where $g\geq 1$) with
scalar field-dependent coupling functions $F^{(g)} (X)$.

As alluded to above, the static, supersymmetric black hole is
constructed as a solitonic interpolation
between two $N=2$ supersymmetric groundstates.
The near-horizon solution can be specified by imposing 
full $N=2$ supersymmetry on the bosonic quantities. A careful
analysis \cite{CDWM} 
of the resulting restrictions on the bosonic background shows 
that the $X^I$ and ${\hat A}$ must be  constant at the horizon.
The near-horizon space-time geometry is determined
to  be of the Bertotti-Robinson type, with $T^{01\,ij} = -i T^{23\,ij}
= 2\,\varepsilon^{ij} \,\exp[-{\cal K}/2]\, \bar Z^{-1}$, while all
other components of $T^{ab\,ij}$ 
vanish. Therefore we have $\hat A =-  64\,\exp[-{\cal K}]\,\bar
Z^{-2}$. 

The requirement of $N=2$ supersymmetry at the horizon does not by itself
fix the actual values of the constants $X^I$.  To do so,
we have to invoke the so-called fixed-point behaviour \cite{FerKalStr}
for the scalar fields $X^I$
at the horizon, according to which, regardless of 
their values at spatial infinity, the $X^I$ (or rather, ratios of the
$X^I$)  
evolve towards the horizon and take values that are determined in
terms of the charges carried by the black hole. This fixed-point
behaviour has been established in the absence of higher-derivative
interactions. Assuming that such a 
behaviour holds in the presence of these interactions, one can
generally argue on the basis of electric/magnetic duality that the
symplectic vectors   
$V$ and $(p^I,q_J)$ must be proportional, which in many cases suffices
to explicitly evaluate the $X^I$ in terms of the charges. However, the
$X^I$ are defined projectively, so that they are only determined modulo an
uniform complex constant. This is one of the reasons why it is
convenient to introduce rescaled variables $Y^I = 
e^{{\cal K}/2}
\bar{Z} X^I$ and $\Upsilon = e^{ {\cal K} } 
\bar{Z}^2 \hat{A} = -64$. Using the homogeneity property of $F$
mentioned earlier, it follows that 
the relation between moduli and charges now reads 
$Y^I - \bar{Y}^I = ip^I$ and $F_I(Y,\Upsilon) - \bar{F}_I(\bar{Y}, 
\bar{\Upsilon}) = i q_I$.  These equations determine the values of
the rescaled fields $Y^I$ in terms of the black hole
charges. Furthermore it follows from (\ref{z}) that $|Z|^2 =
p^I F_I(Y,\Upsilon) - q_I Y^I$, which determines the value of $|Z|$
in terms of $(p^I, q_J)$. 

The entropy of the static black hole solution described above
can now be computed from (\ref{graventro}), using (\ref{Paction}).  
The result takes the remarkably concise form
\be
{\cal S} = \pi \left[ |Z|^2 - 256\, \mbox{Im} \,F_{\hat A} \right] \;.
\label{entropia}
\ee
The first term denotes the Bekenstein-Hawking entropy contribution,
whereas the second term is due to Wald's modification of the definition of
the entropy in the presence of higher-derivative interactions (the
actual contribution originates from $R\, T^2$-terms). 
Note that when switching on the higher-derivative interactions,  the
value of $|Z|$ changes and hence 
also the horizon area changes.  There are thus 
two ways in which the black hole entropy is modified, 
namely by a change of the near-horizon geometry and by an explicit
deviation from the area law.
The entropy (\ref{entropia}) is now entirely determined
in terms of the charges carried by the black hole. 

With these results one can now consider the effective field theory
corresponding to type-IIA string theory compactified on a Calabi-Yau
threefold, in the limit where the volume of the Calabi-Yau threefold
is taken to be large, and compare the result for the black hole
entropy with the results of \cite{MSWV} obtained from the counting of
micro states for the very same objects in the same limit. The
associated homogenous function
$F(Y,\Upsilon)$ is given by (with $I = 0, \dots, n$ and  $A = 1,
\dots, n$) 
\be
F(Y,\Upsilon) 
= \frac{D_{ABC} Y^A Y^B Y^C}{Y^0} + 
d_{A}\, \frac{Y^A}{Y^0} \; \Upsilon\;,
\ee
where the constants $D_{ABC}$ and $d_A$ are related to the
intersection numbers $C_{ABC}$ of the four-cycles and the second
Chern-class number 
$c_{2A}$ of the Calabi-Yau manifold by $D_{ABC} = - \ft16 C_{ABC}$ and
$d_{A} = - \ft{1}{24} \, \ft{1}{64}\; c_{2A}$. 
The Lagrangian (\ref{Paction}) associated with 
this homogenous function thus contains a term proportional to 
$c_{2A} \, {\rm Im }\, z^A \,C_{\mu \nu \rho \sigma}^2$, where 
$z^A = {Y^A}/{Y^0}$.
The $Y^I$ can now be solved for black holes with $p^0 =
0$. Substituting the result into the formula (\ref{entropia}) for the  
macroscopic entropy, one finds \cite{CDWM}, 
\beqa
{\cal S}= 
2 \pi \sqrt{\ft16 \, |{\hat q}_0|  (C_{ABC} \,p^A p^B p^C + 
{c}_{2A} \, p^A) }\;\;,
\label{entrot2}
\eeqa
where ${\hat q}_0 = q_0 + \ft{1}{12} D^{AB} q_A q_B$ with $D_{AB} =
D_{ABC} p^C$ and $D_{AB} D^{BC} = \delta_A^C$.  The expression
(\ref{entrot2}) for the macroscopic entropy is 
in exact agreement with the microscopic entropy formula computed in
\cite{MSWV} via state counting.

\vspace{.5cm}


\end{document}